\documentclass[twocolumn]{revtex4}

\usepackage{graphicx}
\usepackage{epsf}
\usepackage{amsmath,amssymb}
\usepackage{color}
\newcommand{\bvec}{\boldsymbol}

\begin{document}
\preprint{KUNS-2751}
\title{Isoscalar dipole excitations in $^{16}$O}
\author{Yoshiko Kanada-En'yo}
\affiliation{Department of Physics, Kyoto University, Kyoto 606-8502, Japan}
\author{Yuki Shikata}
\affiliation{Department of Physics, Kyoto University, Kyoto 606-8502, Japan}
\begin{abstract}
Isoscalar (IS) monopole and dipole
excitations in $^{16}$O were investigated by the method of  shifted basis antisymmetrized molecular dynamics 
combined with the generator coordinate method. 
Significant strengths of the IS monopole and dipole transitions were obtained in the low-energy region 
below the giant resonances.
In addition to the compressive mode, which mainly contributes to the high-energy strengths for the IS dipole giant resonance, we obtained a variety of low-energy dipole modes such as  
the vortical dipole mode in the $1^-_1$ state of the vibrating tetrahedral $4\alpha$ and 
the $^{12}$C+$\alpha$ cluster structure in the $1^-_2$ state.
The $1^-_1$ state contributes to 
the significant low-energy strength of the IS dipole transition as 5\% of the energy-weighted sum rule,
which describes well the experimental data observed by the $\alpha$ inelastic scattering.
\end{abstract}
\maketitle

\section{Introduction}
In the past decades, low-energy  monopole and dipole excitations have been attracting
great interests
(see, for example, reviews in 
Refs.~\cite{Harakeh-textbook,Paar:2007bk,aumann-rev,Savran:2013bha,Bracco:2015hca} and references therein). 
A central issue is possible appearance of new excitation modes decoupled from
collective vibration modes corresponding the giant resonances (GR). In 
experiments with $\alpha$ inelastic scattering extensively performed 
for study of isoscalar (IS) monopole and dipole excitations, 
significant low-energy strengths with the fraction of several percentages 
of the energy weighted sum rule have been observed in various stable nuclei 
such as $^{16}$O, $^{40}$Ca, and $^{208}$Pb  
\cite{Harakeh:1981zz,Decowski:1981pcz,Poelhekken:1992gvp}. 
The questions to be answered are what is the origin of these IS low-energy dipole (LED) strengths and 
how the dipole modes come down to the energy much lower  
than the IS giant dipole resonances (GDR).

In order to understand the IS LED strengths,
the vortical dipole (VD) mode (called also the torus or toroidal mode) has been studied
firstly with hydrodynamical models \cite{semenko81,Ravenhall:1987thb}, 
and later with microscopic approaches \cite{Paar:2007bk,Vretenar:2001te,Ryezayeva:2002zz,Papakonstantinou:2010ja,Kvasil:2011yk,Repko:2012rj,Kvasil:2013yca,Nesterenko:2016qiw,Nesterenko:2017rcc}.
The VD mode is characterized by the nuclear vorticity and has a unique feature different from 
the standard IS dipole mode so-called compressive dipole (CD) in the IS GDR. Since the 
nuclear density is conserved in the VD mode, its energy can be lower than the IS GDR involving compression of nuclear density. 
As a measure of the nuclear vorticity in the dipole excitations, 
the toroidal dipole (TD) operator 
has been introduced \cite{semenko81,Dubovik75}. The TD operator is given by the rotational component (a curl term) of the 
transition current density and the counter part of the compressive dipole (CD) operator with the 
irrotational component (a divergence term) of the transition current density, and has been proved to be 
a good probe for the low-energy VD mode \cite{Kvasil:2011yk}.
% rather than the VD operator introduced by \cite{Ravenhall:1987thb}.

In light nuclei, also cluster states may contribute to the low-energy IS monopole (IS0) and dipole (IS1) transition strengths
because the IS0 and IS1 operators contain higher order $r^{\lambda+2}$ terms 
and can excite not only the compressive vibration modes but also the inter-cluster motion 
in the cluster states as pointed out by Yamada {\it et al.} \cite{Yamada:2011ri} and Chiba {\it et al.} \cite{Chiba:2015khu}.
Indeed, the low-energy IS monopole strengths in $^{16}$O have been described well by cluster states with
a semi-microscopic $4\alpha$-cluster model %of the orthogonal condition model (OCM)
\cite{Yamada:2011ri}. It is an important issue to clarify the IS dipole excitations  in $^{16}$O, in particular, the
cluster and vortcal aspects of the low-energy modes. 

Theoretical calculations with cluster models have been performed for $^{16}$O and 
suggested  a variety of cluster structures such as the tetrahedral $4\alpha$ and 
$^{12}$C+$\alpha$ structures \cite{Yamada:2011ri,wheeler37,dennison54,brink70,Suzuki:1976zz,Suzuki:1976zz2,fujiwara80,Libert-Heinemann:1980ktg,bauhoff84,Descouvemont:1987uvu,Descouvemont:1991zz,Descouvemont:1993zza,Fukatsu92,Funaki:2008gb,Funaki:2010px,Kanada-En'yo:2013dma,Horiuchi:2014yua}. However,
there have been no microscopic calculation that 
successfully describes the energy spectra of $^{16}$O. Recently, we applied a microscopic model of the antisymmetrized 
molecular dynamics (AMD) \cite{KanadaEnyo:1995tb,KanadaEnyo:1995ir,KanadaEn'yo:2001qw,KanadaEn'yo:2012bj}
to $^{16}$O, and obtained reasonable reproduction of the energy spectra of $^{16}$O such as 
$0^+_2$, $2^+_1$, $4^+_1$, $1^-_2$, and $3^-_2$  states in the positive- and negative-parity bands with the 
$^{12}$C+$\alpha$ structure and 
$3^-_1$ and $4^+_2$ states in the ground band with the tetrahedral $4\alpha$ structure \cite{Kanada-En'yo:2013dma,Kanada-Enyo:2017ers}. 

Our aim is to investigate the IS dipole excitations in $^{16}$O. 
Main interest are properties of the IS LED modes such as the cluster and vortical aspects. 
For this aim, we apply the method of the shifted basis AMD (sAMD) \cite{Kanada-Enyo:2015knx,Kanada-Enyo:2015vwc,Kimura:2016heo} combined with the cluster generator coordinate method (GCM). 
The sAMD+GCM has been recently constructed to 
describe both the single-particle excitation and large amplitude cluster mode. 
This method has been applied to $^{12}$C to discuss the cluster,  vortical, and compressive IS dipole modes, and 
proved to be a powerful approach 
for the IS monopole and dipole excitations in a wide energy range including the low-energy states and high-energy GRs. 
\cite{Kanada-Enyo:2015vwc,Kanada-Enyo:2017fps}.

In our previous work of $^{16}$O\cite{Kanada-Enyo:2017ers},  
we have investigated the cluster states 
with variation after spin-parity projections (VAP) \cite{Kanada-Enyo:1998onp} combined with the $^{12}$C+$\alpha$-cluster 
GCM, which we called the VAP+GCM, but not the IS GDR because the sAMD bases have not been adopted 
in the previous work. The great advantages of the present sAMD+GCM are
that it describes both the low-energy cluster state and the GDR in a unified framework owing to inclusion of one-particle and 
one-hole (1p-1h) excitations in the sAMD bases, and is suitable to discuss details of the IS dipole excitations. 
In this paper, we show the IS monopole and dipole strength functions in $^{16}$O in a wide energy range 
covering the low-lying vortical and cluster modes, and also the high-energy compressive vibration modes of the GRs.
For detailed analysis of the monopole and dipole transitions, we calculate the form factors and transition densities and 
compare them with experimental data measured by the electron scattering. We discuss 
the vortical and cluster aspectes of the IS LED states and clarify properties of 
the IS dipole excitations. 

The paper is organized as follows. 
The formulation of the sAMD+GCM for $^{16}$O is explained in Sec.~\ref{sec:framework}.
Section  \ref{sec:results} shows 
the calculated results and discusses the properties of the IS monopole and dipole modes.
%In addition, cluster model analyses are presented in Sec.~\ref{sec:discussions}.
Finally, the paper is summarized in section \ref{sec:summary}.
In appendix sections, the definitions of the transition operators, densities, and strengths are given.

\section{Formulation}\label{sec:framework}
In order to calculate the IS monopole and dipole excitations in $^{16}$O,
we combine the sAMD with the previous VAP+GCM model \cite{Kanada-Enyo:2017ers}. 
Namely, we prepare the sAMD wave functions and combine them with the 
the basis wave functions adopted 
in the previous VAP+GCM calculation. 
We call the present calculation ``sAMD+GCM''.
In this section, we explain the framework and procedure of the present calculations of $^{16}$O.
For details of the VAP+GCM and the sAMD, the reader is 
referred to Refs.~\cite{Kanada-Enyo:2015vwc,Kanada-Enyo:2017fps,Kanada-Enyo:2017ers,Kanada-Enyo:2017uzz}
and references therein. 

\subsection{VAP+GCM with AMD wave functions}
An AMD wave function is given by a Slater determinant of single-particle Gaussian wave functions,
\begin{eqnarray}
 \Phi_{\rm AMD}({\bvec{Z}}) &=& \frac{1}{\sqrt{A!}} {\cal{A}} \{
  \varphi_1,\varphi_2,...,\varphi_A \},\label{eq:slater}\\
 \varphi_i&=& \phi_{{\bvec{X}}_i}\chi_i\tau_i,\\
 \phi_{{\bvec{X}}_i}({\bvec{r}}_j) & = &  \left(\frac{2\nu}{\pi}\right)^{4/3}
\exp\bigl\{-\nu({\bvec{r}}_j-\bvec{X}_i)^2\bigr\},
\label{eq:spatial}\\
 \chi_i &=& (\frac{1}{2}+\xi_i)\chi_{\uparrow}
 + (\frac{1}{2}-\xi_i)\chi_{\downarrow}.
\end{eqnarray}
where  ${\cal{A}}$ is the antisymmetrizer, 
$\phi_{{\bvec{X}}_i}$, $\chi_i$, and $\tau_i$ are the spatial,  spin, and isospin functions of 
the $i$th single-particle wave function, respectively. 
The isospin part is fixed to be up (proton) or down (neutron). 
$\nu$ is the width parameter, which is fixed to be $\nu=0.19$ fm$^{-2}$ 
used in the previous calculation.
The condition $\sum_{i=1,\ldots,A} \bvec{X}_i/A=0$ is always kept and 
the contribution of the center of mass motion is exactly removed from the total system. 
The AMD wave function is specified by the set of variational  
parameters ${\bvec{Z}}\equiv 
\{{\bvec{X}}_1,\ldots, {\bvec{X}}_A,\xi_1,\ldots,\xi_A \}$ for the centroids of single-nucleon Gaussian wave packets 
and nucleon-spin orientations, which are determined by the energy variation. 

It should be stressed that,   in the AMD model, the existence of any clusters is not {\it a priori} assumed
because Gaussian centroids, ${\bvec{X}}_1,\ldots,{\bvec{X}}_A$, of
all single-nucleon wave packets are independently treated as variational parameters. 
Nevertheless, the model wave function can
describe various cluster wave functions, and also shell-model wave functions
because of the antisymmetrization of Gaussian wave packets.
%It should be also commented that similar wave function is used in the FMD framework
% \cite{Feldmeier:1994he,Neff:2002nu}.

%For a $J^\pi$ state, the spin-parity projected AMD wave function is used:
%\begin{equation}
%\Phi^{J\pi}(\bvec{Z})=P^{J\pi}_{MK}\Phi_{\rm AMD}(\bvec{Z}),
%\end{equation}
%where $P^{J\pi}_{MK}$ is the spin-parity projection operator.
To obtain the AMD wave function optimized for the $J^\pi$ state, 
the VAP is performed with respect to the variation of $\bvec{Z}$ by  
\begin{equation}
\delta\frac{\langle \Phi|H|\Phi \rangle}{\langle \Phi|\Phi \rangle}=0,
\end{equation}
for the  $J^\pi$-projected AMD wave function $\Phi=P^{J\pi}_{MK}\Phi_\textrm{AMD}(\bvec{Z})$,
where $P^{J\pi}_{MK}$ is the spin-parity projection operator.
For the AMD wave function  $\Phi^{^{16}\textrm{O}}_\textrm{AMD}(\bvec{Z})$ of $^{16}$O, we perform
the VAP with
$J^\pi_k=0^+_{1,2}$, 
$2^+_1$, $4^+_{1,2}$, $1^-_{1}$, $2^-_{1}$, $3^-_{1}$, and 
$5^-_1$, and obtain nine configurations of $\Phi^{^{16}\textrm{O}}_\textrm{AMD}(\bvec{Z}^{\rm opt}_{\beta})$ with 
the parameters $\bvec{Z}^{\rm opt}_{\beta}$ optimized for each $\beta=J^\pi_k$ state.
In the simple VAP calculation, we superpose the nine configurations.

In the GCM calculation, we adopt the $^{12}$C+$\alpha$ cluster wave functions, where 
the angular momentum projection and internal excitations of the sub system $^{12}$C-cluster are considered.
We first perform the VAP calculation of the subsystem $^{12}$C for three states $^{12}$C($0^+_1)$,  
$^{12}$C($0^+_2)$,  and  $^{12}$C($1^-_1)$. Using the obtained $^{12}$C-cluster wave functions, 
the $^{12}$C+$\alpha$ wave function is constructed as done in Ref.~\cite{Kanada-Enyo:2017ers}. 
The relative distance $d$ between $^{12}$C and $\alpha$ clusters
is treated as a generator coordinate. The angular-momentum projection of the subsystem $^{12}$C is also practically 
performed by taking into account rotation of the $^{12}$C-cluster.

\subsection{sAMD+GCM:\ combination of sAMD with VAP+GCM}
In addition to the VAP and $^{12}$C+$\alpha$ wave functions, the sAMD wave functions are also superposed
to describe 1p-1h excitations on the ground state. 
Starting from the ground state wave function 
$\Phi^{^{16}\textrm{O}}_\textrm{AMD}(\bvec{Z}^{\rm opt}_{\beta=0^+_1})$ obtained by the VAP, 
we consider small variations of single-particle wave functions 
by shifting the Gaussian centroid 
of each single-particle wave function, 
${\bvec{X}}_i\rightarrow {\bvec{X}}_i+\epsilon{\bvec{e}}_\sigma$ (the spatial position parameters),  
of $\bvec{Z}^{\rm opt}_{\beta=0^+_1}$ in the AMD wave function. 
Here $\epsilon$ is an enough small constant, 
${\bvec{e}}_\sigma$ ($\sigma=1,\ldots,8$) are unit vectors for 8 directions.
Spin non-flip and flip states and recoil effects are taken into account as 
explained in Ref.~\cite{Kanada-Enyo:2015vwc}. 
Consequently, 
totally $16A=256$ bases
of the shifted AMD wave functions are superposed in addition to 
the VAP and $^{12}$C+$\alpha$ wave functions 
in the sAMD+GCM calculation of $0^+$ and $1^-$ states. 

In the present sAMD+GCM calculation, we use the $\Phi^{^{12}\textrm{C}}_\textrm{AMD}(\bvec{Z}^{\rm opt}_{\beta=0^+_1})$+$\alpha$ configuration 
with the 
inter-cluster distances of 
 $d=\{1.2, 2.4, \ldots, 7.2$~fm\} and $\Phi^{^{12}\textrm{C}}_\textrm{AMD}(\bvec{Z}^{\rm opt}_{\beta=0^+_2,1^-_1})$+$\alpha$  
configurations with 
 $d=\{1.2, 2.4, \ldots, 4.8$~fm\} to save the computational cost.
($\Phi^{^{12}\textrm{C}}_\textrm{AMD}(\bvec{Z}^{\rm opt}_{\beta=0^+_1,0^+_2,1^-_1})$+$\alpha$ with 
$d=\{1.2, 2.4, \ldots, 8.4$~fm\}  are used in Ref.~\cite{Kanada-En'yo:2013dma}, and 
$\Phi^{^{12}\textrm{C}}_\textrm{AMD}(\bvec{Z}^{\rm opt}_{\beta=0^+_1,0^+_2})$+$\alpha$ with 
$d=\{1.2, 2.4, \ldots, 8.4$~fm\}  and 
$\Phi^{^{12}\textrm{C}}_\textrm{AMD}(\bvec{Z}^{\rm opt}_{\beta=1^-_1})$+$\alpha$ with 
$d=\{1.2, 2.4, \ldots, 4.8$~fm\}  are used in Ref.~\cite{Kanada-Enyo:2017ers}.)

The IS0 and IS1 transition strengths are calculated with
the $0^+$ and $1^-$ states obtained by the sAMD+GCM.
The form factors and transition densities are also calculated with these operators.  
As for the IS dipole  excitations, transition strengths of the CD and TD operators 
are also calculated. 
%The former (CD) is in principle equivalent to the
%standard IS1 operator, and the latter (TD) is the operator sensitive to the 
%nuclear vorticity. 
The definitions of the operators, matrix elements, strengths, form factors, and 
transition densities are given in appendixes.

\section{Results}\label{sec:results}

\subsection{Structure properties of low-energy levels of  $0^+$ and $1^-$ states}
The sAMD+GCM result of the 
binding energy, root-mean-square (rms) matter radii, and excitation energies of low-lying $0^+$ states 
are listed in Table \ref{tab:structure-l0p}, and those of the $1^-_1$ and $1^-_2$ states are shown in 
Table \ref{tab:structure-l1n}. 
For comparison, values calculated with the VAP (without the $^{12}$C+$\alpha$ nor 
sAMD bases) and those of the VAP+GCM (without the sAMD bases) are also shown in the tables. 
These corresponds to the VAP and VAP+GCM calculations presented in the previous paper 
\cite{Kanada-Enyo:2017ers}.

Various cluster states are obtained in the excited $0^+$ levels in $E\lesssim 20$ MeV.
Compared the sAMD+GCM and VAP+GCM, 
there is no essential difference  between the two calculations for these states, 
because the developed cluster states are dominantly 
contributed by the GCM  bases but not by the sAMD bases.
It is not the case for the ground state, but 
the sAMD+GCM obtains 2 MeV energy gain of the $0^+_1$ state 
compared with the VAP+GCM meaning that the sAMD bases efficiently improve the ground state correlations.
Because of this additional energy gain of the ground state, the relative
energy position of the excited $0^+$ states are raised up by about 2 MeV in the sAMD+GCM.
As a result, the agreement with the experimental energy spectra  in the sAMD+GCM is not 
as good as the VAP+GCM, but it is much better than the preceding 
microscopic cluster model calculations. % and may be allowable.
%The energy spectra could be improved by inclusion of 
%sAMD bases for the subsystem $^{12}$C-cluster wave functions but it is beyond the scope of the present 
%paper. 
We note that, the calculated fourth $0^+$ state with the $^{12}$C$(2^+_1)$+$\alpha$ cluster structure 
should be assigned to the experimental $0^+_3$ state, 
because the sAMD+GCM and VAP+GCM calculations eventually give the opposite ordering of the $0^+_3$ and $0^+_4$ states
as discussed in the previous paper.
%properties of the $0^+_3$ is described well by the $^{12}$C$(2^+_1)$+$\alpha$ clustering 
%as discussed with the OCM (semi-microscopic) calculations \cite{Suzuki,Yamada,Funaki}
%though

In the calculated $1^-$ levels, the $1^-_1$ and $1^-_2$ states are obtained in $E < 15$ MeV.
The higher state ($1^-_2$) is the well developed cluster state and 
regarded as the band-head state of the $K^\pi=0^-$ $^{12}$C+$\alpha$ band, which is 
the parity doublet of the $K^\pi=0^+_2$ $^{12}$C+$\alpha$  band built on the band-head $0^+_2$ state.
The lower state ($1^-_1$) has the small rms radius comparable to that of the ground state and shows 
less prominent cluster structure than the $1^-_2$ and $0^+_{2,3,4,5}$ states. 
Comparing with the VAP+GCM, 
the sAMD+GCM gives the smaller radius of the $1^-_1$ state. Moreover, 
the excitation energy of the $1^-_1$ state is almost same between the  sAMD+GCM and VAP+GCM calculations
indicating that the sAMD bases describe additional correlations contributing the size shrinkage and 
the 2 MeV energy gain comparable to that of the ground state.
%This is consistent with the picture 
%for the $1^-_1$ as the vibration mode on the tetrahedral $4\alpha$ structure of the ground state, 
%which have been discussed with the algebraic cluster approach \cite{}.

\begin{table}[ht]
\caption{Properties of $0^+$ states; the binding energy (B.E.), excitation energies ($E_x$), 
rms matter radii ($R$), 
and the IS0 matrix elements ($M(E0)$). The present result (sAMD+GCM) and the 
VAP and VAP+GCM values from Ref.~\cite{Kanada-Enyo:2017ers} are shown compared 
with the experimental data \cite{Tilley:1993zz}.
The experimental value of the rms radius of the ground state
is deduced from the experimental charge radius measured by the electron scattering\cite{Angeli2013}.
 \label{tab:structure-l0p}
}
\begin{center}
\begin{tabular}{cccccccccc}
\hline
		&	VAP	&	VAP	&	sAMD&	exp	\\
		& &	+GCM	&	+GCM			\\
$	\textrm{B.E. (MeV)}	$&	123.0 	&	123.5 	&	125.6 	&	127.62 	\\
$		$&		&		&		&		\\
$	E_x(0^+_2) \ \textrm{(MeV)}	$&	13.1 	&	9.7 	&	11.6 	&	6.05	\\
$	E_x(0^+_3) \ \textrm{(MeV)}	$&		&	15.3 	&	18.6 	&	12.05	\\
$	E_x(0^+_4) \ \textrm{(MeV)}	$&		&	13.6 	&	15.5 	&	13.6	\\
$	E_x(0^+_5) \ \textrm{(MeV)}	$&		&	18.3 	&	20.6 	&	14.01	\\
$		$&		&		&		&		\\										
$	R(0^+_1)\ \textrm{(fm)}	$&	2.69 	&	2.73	&	2.72 	&	2.55	\\
$	R(0^+_2)\ \textrm{(fm)}	$&	2.96 	&	3.29 	&	3.16 	&		\\
$	R(0^+_3)\ \textrm{(fm)}	$&		&	3.53 	&	3.45 	&		\\
$	R(0^+_4)\ \textrm{(fm)}	$&		&	3.64 	&	3.21 	&		\\
$	R(0^+_5)\ \textrm{(fm)}	$&		&	3.53 	&	3.36 	&		\\
$		$&		&		&		&		\\
%$	\langle 0^+_1|\bvec{S}^2_p |0^+_1 \rangle	$&	0.07 							\\
%$	\langle 0^+_2|\bvec{S}^2_p|0^+_2 \rangle	$&	0.71 							\\
%$		$&		&		&		&		\\
$	M(E0;0^+_1\to0^+_2)	$ ($e$\ fm$^2$) &	1.8 	&	3.5 	&	3.8 	&	3.55(0.21)	\\
$	M(E0;0^+_1\to0^+_3)	$ ($e$\ fm$^2$) &		&	3.3 	&	3.9 	&	4.03(0.09)	\\
$	M(E0;0^+_1\to0^+_4)	$ ($e$\ fm$^2$)&		&	4.1 	&	4.1 	&		\\
$	M(E0;0^+_1\to0^+_5)	$ ($e$\ fm$^2$)&		&	3.0 	&	3.2 	&	3.3(0.7)	\\
\hline	
\end{tabular}
\end{center}
\end{table}

\begin{table}[ht]
\caption{Properties of the $1^-_1$ and $1^-_2$ states, 
excitation energies, rms radii, 
the IS1 strengths, and the EWSR ratio $P_\textrm{IS1}$ of the energy-weighted IS1 strengths.
The present result of the sAMD+GCM and those of the 
VAP and VAP+GCM calculations from Ref.~\cite{Kanada-Enyo:2017ers} are shown compared 
with the experimental data \cite{Tilley:1993zz}.
The experimental data of the EWSR ratio 	$P_\textrm{IS1}(1^-_1)$
is the value from Ref.~\cite{Harakeh:1981zz} of $\alpha$ inelastic scattering analysis.
 \label{tab:structure-l1n}
}
\begin{center}
\begin{tabular}{cccccccccc}
\hline
		&	VAP	&	VAP	&	sAMD&	exp	\\
		& &	+GCM	&	+GCM			\\
$	E_x(1^-_1)\ \textrm{(MeV)}	$&	10.3 	&	9.4 	&	9.6 	&	7.12	\\
$	E_x(1^-_2)\ \textrm{(MeV)}	$&	17.0 	&	12.1 	&	14.4 	&	9.59	\\
$		$&		&		&		&		\\
$	R(1^-_1)\ \textrm{(fm)}	$&	2.76 	&	2.87 	&	2.80 	&		\\
$	R(1^-_2)\ \textrm{(fm)}	$&	2.96 	&	3.58 	&	3.37 	&		\\
$		$&								\\
$	B(\textrm{IS1};0^+_1\to 1^-_1)	$&	124.5 	&	165.5 	&	169.8 	&		\\
$	P_\textrm{IS1}(1^-_1)	$&	0.042 	&	0.048 	&	0.051 	&	0.42$^a$	\\
$	B(\textrm{IS1};0^+_1\to 1^-_2)	$&	7.9 	&	2.9 	&	10.2 	&		\\
$	P_\textrm{IS1}(1^-_2)	$&	0.0044 	&	0.0011 	&	0.0045 	&		\\
\hline	
\end{tabular}
\end{center}
\end{table}

\subsection{Cluster structures of low-lying states}
Cluster aspects of the low-lying states have been investigated in Ref.~\cite{Kanada-Enyo:2017ers}. 
We here briefly review the cluster structures of the $0^+_1$, $0^+_2$, $1^-_1$, and $1^-_2$ states
following the discussions in the previous paper 
based on the analysis of  the intrinsic wave functions, 
$\Phi^{^{16}\textrm{O}}_\textrm{AMD}(\bvec{Z}^{\rm opt}_{\beta=J^\pi_k})$, obtained by the VAP calculation.

%%%%%%%%%%%%%%%%%%%%%%%%%%%%%%
\begin{figure}[!h]
\begin{center}
\includegraphics[width=8cm]{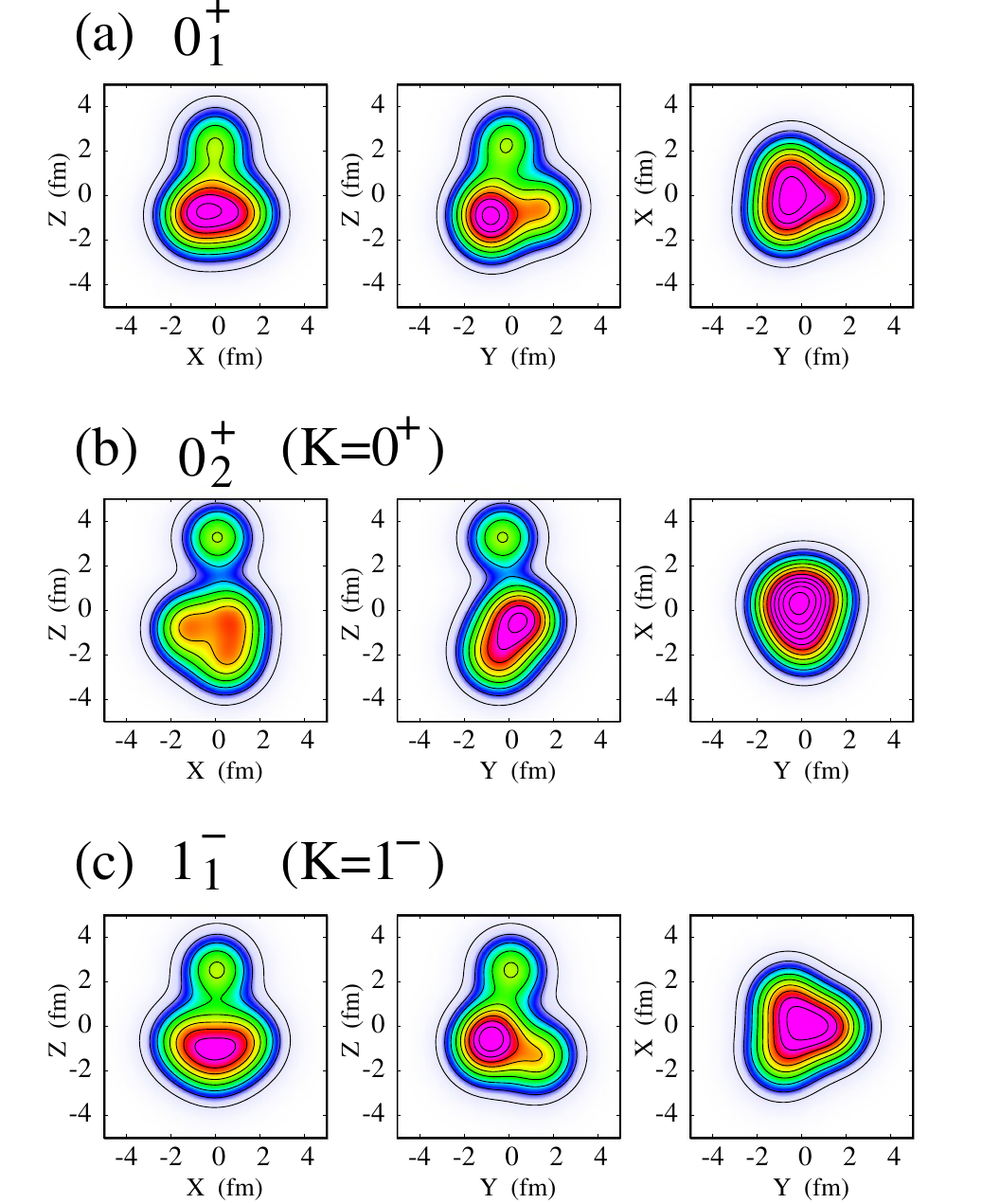} 	
\end{center}
%\vspace{0.5cm}
  \caption{(color online)  
Density distributions in the intrinsic states obtained by the 
VAP for the $0^+_1$, $0^+_2$, and $1^-_1$ states.
The densities integrated along the $Y$, $X$, and $Z$ axes are plotted on the (left) $X$-$Z$, 
(middle) $Y$-$Z$, and (right) $X$-$Y$ planes, respectively. 
Figures corresponds to those of Ref.~\cite{Kanada-Enyo:2017ers}, 
but are reconstructed from the wave functions.
\label{fig:dense}}
\end{figure}
%%%%%%%%%%%%%%%%%%%%%%%%%%%%%

Figure \ref{fig:dense} shows the intrinsic density distribution of the  $0^+_1$,  $0^+_2$, and  $1^-_2$ states. 
The $0^+_1$ state shows the tetrahedral $4\alpha$ cluster structure, in which 
 three $\alpha$s form the triangle  shape on the $X$-$Y$ plane 
and the last $\alpha$ cluster is sitting on the $Z$(vertical) axis (Fig.~\ref{fig:dense}(a)). 
Its cluster development is not so remarkable as seen in the compact density distribution.
The $1^-_1$ state also has a tetrahedral $4\alpha$ clustering with a compact density distribution 
similar to the  $0^+_1$ state, but the orientation of the triangle $3\alpha$ part is somewhat tilted from
the $0^+_1$.  This tilting motion of the triangle $3\alpha$ 
produces the dipole excitation with $K^\pi=1^-$ in the $1^-_1$. This mode is similar to the 
vibration mode of the tetrahedral $4\alpha$ discussed 
by the algebraic $4\alpha$ cluster model \cite{Bijker:2014tka,Bijker:2016bpb}. However, 
the $0^+_1$ and $1^-_1$ states obtained in the present calculation are not the equilateral tetrahedral states 
but the prolately deformed one with the $3\alpha+\alpha$ configuration and contain the $\alpha$ breaking component.  

The $0^+_2$ state has the developed $^{12}$C+$\alpha$ cluster structure, in which 4$\alpha$ clusters are arranged
in a planar-like configuration. Because of the remarkably developed $^{12}$C+$\alpha$ clustering, 
the $0^+_2$ state shows a largely deformed intrinsic density compared with the $0^+_1$.  
The developed $^{12}$C+$\alpha$ clustering 
constructs the $K^\pi=0^+$ band and the parity-partner $K^\pi=0^-$ band starting from the 
the band-head $1^-_2$ state.

We should note that, even though the $0^+_1$,  $0^+_2$, $1^-_1$, and $1^-_2$
show the formation of four $\alpha$ clusters, the clusters 
are not necessarily the ideal $\alpha$ clusters with the $(0s)^4$ configuration but contain the $\alpha$-cluster breaking
because of the spin-orbit interaction.
We can evaluate the $\alpha$-cluster breaking component from
the expectation value of the squared proton spin $\langle \bvec{S}_p^2 \rangle$ because it measures the $S=1$ mixing induced by the 
the $\alpha$ breaking. 
The values calculated with the VAP
are $\langle \bvec{S}_p^2 \rangle=0.07$, 0.71, 0.35, and 0.79 for the $0^+_1$,  $0^+_2$, $1^-_1$, and $1^-_2$, respectively, 
indicating the slight breaking in the $0^+_1$ and 
the significant $\alpha$ breaking in the $0^+_2$, $1^-_1$, and $1^-_2$.

It should be also commented that, 
these VAP configurations couple with other configurations such as the $^{12}$C-cluster rotation and
1p-1h excitations in the sAMD+GCM calculation, 
but they still give significant contributions and roughly describe main properties of 
the $0^+_1$, $0^+_2$, $1^-_1$, and $1^-_2$ states. 
 
\subsection{Transition strengths}

%The definitions of the ISM and ISD operators and transition strengths are explained in appendix. 
The calculated IS0 and IS1 transition strengths to the $0^+_{2,3,4,5}$ and $1^-_{1,2}$ states
are  listed in Tables \ref{tab:structure-l0p} and \ref{tab:structure-l1n}. Here 
the strengths $B(E0)=B(\textrm{IS0})/4$ are compared with the experimental data. 
%The calculated results for these low-lying states are again 
%consistent between the VAP+GCM and sAMD+GCM calculations. 
The observed  $B(E0)$ of the $0^+_2$, $0^+_3$, and $0^+_5$ state 
are reproduced well by the sAMD+GCM calculation. 
In the dipole excitations, the remarkably large $B(\textrm{IS1})$ is obtained for the $1^-_1$ 
with the energy weighted sum rule ratio of 5\%, where as the much weaker IS1 transition is obtained 
for the $1^-_2$ state in the $^{12}$C+$\alpha$ band.
The relatively weak IS1 transition to the cluster state seems to contradict the naive expectation that 
the compressive operator could excite cluster states, but it is not true the case of the $1^-_2$ state. 
As mentioned previously, the $1^-_2$ state in the $^{12}$C+$\alpha$ band
has the planar-like configuration and shows the different orientation of the triangle $^{12}$C-cluster
from the initial $0^+_1$. Therefore, the $0^+_1$ to $1^-_2$ excitation
involves not only the inter-cluster excitation but also the $^{12}$C-cluster rotation, which can not be directly excited by the 
IS1 operator. 

\begin{figure}[!h]
\begin{center}
\includegraphics[width=8.6cm]{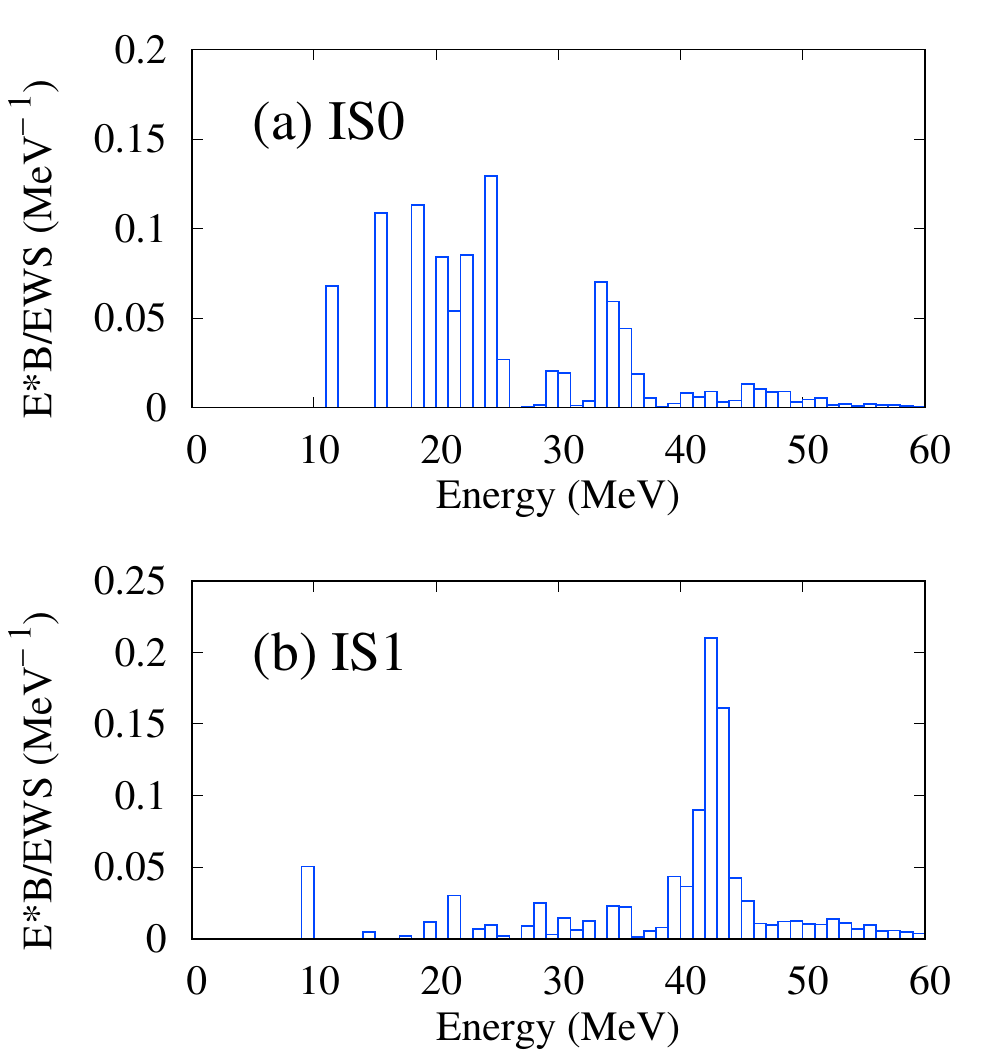} 	
\end{center}
%\vspace{0.5cm}
  \caption{The EWSR ratio of the IS0 and IS1 transition strengths 
calculated with the sAMD+GCM.
\label{fig:s-function}}
\end{figure}

The IS0 and IS1 strength functions up to up to $E=60$ MeV
are shown in Fig.~\ref{fig:s-function}. 
The energy weighted sum rule ratios calculated with the sAMD+GCM are plotted. 
In the IS0 strength function, a large fraction of the strengths are distributed in $E\le 40$ MeV. The cluster states 
significantly contribute to the lower part of the strengths in $E\le 20$ MeV, which are not clearly separated 
from the GMR strengths. 
On the other hand, in the IS1 strength function, the $1^-_1$ state contributes to the significant low-energy strength 
separated from the IS GDR peak around 40 MeV.
% The clear separation indicates 
%existence of the LED mode decoupled from the GDR mode. 

%%%%%%%%%%%%%%%%%%%%%%%%%%%%%%
%%%%%%%%%%%%%%%%%%%%%%%%%%%%%

\subsection{Form factors and transition densities}
\begin{figure}[!h]
\begin{center}
\includegraphics[width=7cm]{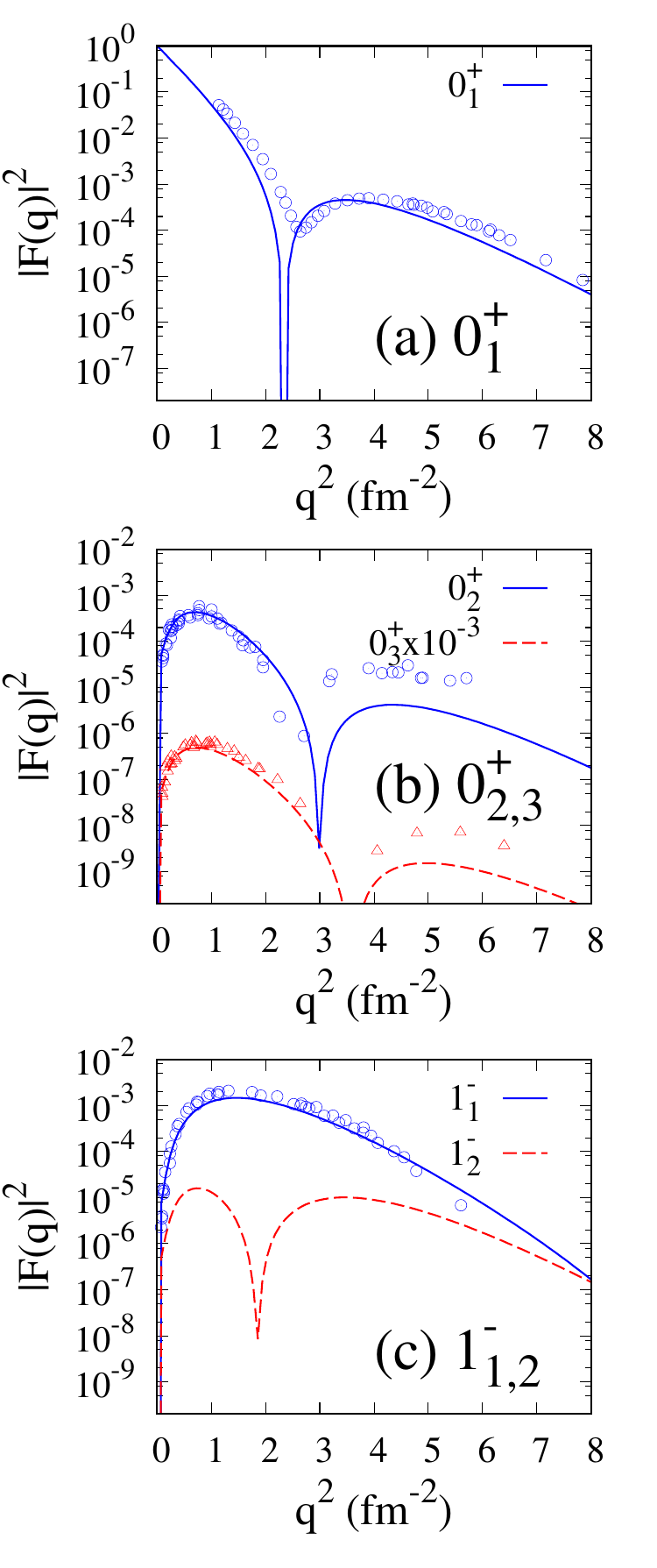} 	
\end{center}
%\vspace{0.5cm}
  \caption{(color online)
The  elastic and inelastic form factors of the IS0 and IS1 transitions
calculated with the sAMD+GCM.
The experimental data are electron scattering form factors 
from Ref.~\cite{Buti:1986zz}.
\label{fig:form}}
\end{figure}

Figure \ref{fig:form} shows the calculated 
elastic and inelastic form factors of the IS0 and IS1 transitions from the ground state 
to the $0^+_{1,2,3}$ and 
$1^-_{1,2}$ states in comparison with experimental data  observed by electron scattering \cite{Buti:1986zz}.

The calculated form factors of the $0^+_{1,2,3}$ states 
are in good agreement with the experimental data 
in the low-momentum region. 
In the shape of the observed inelastic form factors, 
a difference can be seen between the $0^+_2$ and $0^+_3$ states. 
The form factor of the $0^+_2$ drops off at the smaller transfer momentum $q$ 
than the $0^+_3$ reflecting the broader radius dependence of the transition density of the $0^+_2$. 
This trend is qualitatively described in the present calculation and understood by 
the difference in the cluster structures between the 
$0^+_2$ and $0^+_3$ states: the dominant $^{12}$C$(0^+_1)$+$\alpha$ component in the 
$0^+_2$ state and the $^{12}$C$(2^+_1)$+$\alpha$ component in the $0^+_3$ state. 

For the dipole transition to the $1^-_1$,
the magnitude and shape of the experimental form factor are nicely reproduced by the
present calculation. Compared to the  $1^-_1$, 
the calculated IS1 transition to the $1^-_2$ is quite weak.
At the maximum peak, the form factor of the $0^+_1\to 1^-_2$ transition is 
about two orders less than that of  the $0^+_1\to 1^-_1$ transition.  Moreover, the shape of the form factor
is different between the $1^-_1$ and $1^-_2$ states because of the structure difference.
The form factor of the $1^-_2$ in the $^{12}$C+$\alpha$ band 
shows two peak structure with a dip at $q\sim 2$ fm$^{-2}$, which can not be seen in the
form factor of the  $1^-_1$ state with the compact tetrahedral $4\alpha$.

\begin{figure}[!h]
\begin{center}
\includegraphics[width=8cm]{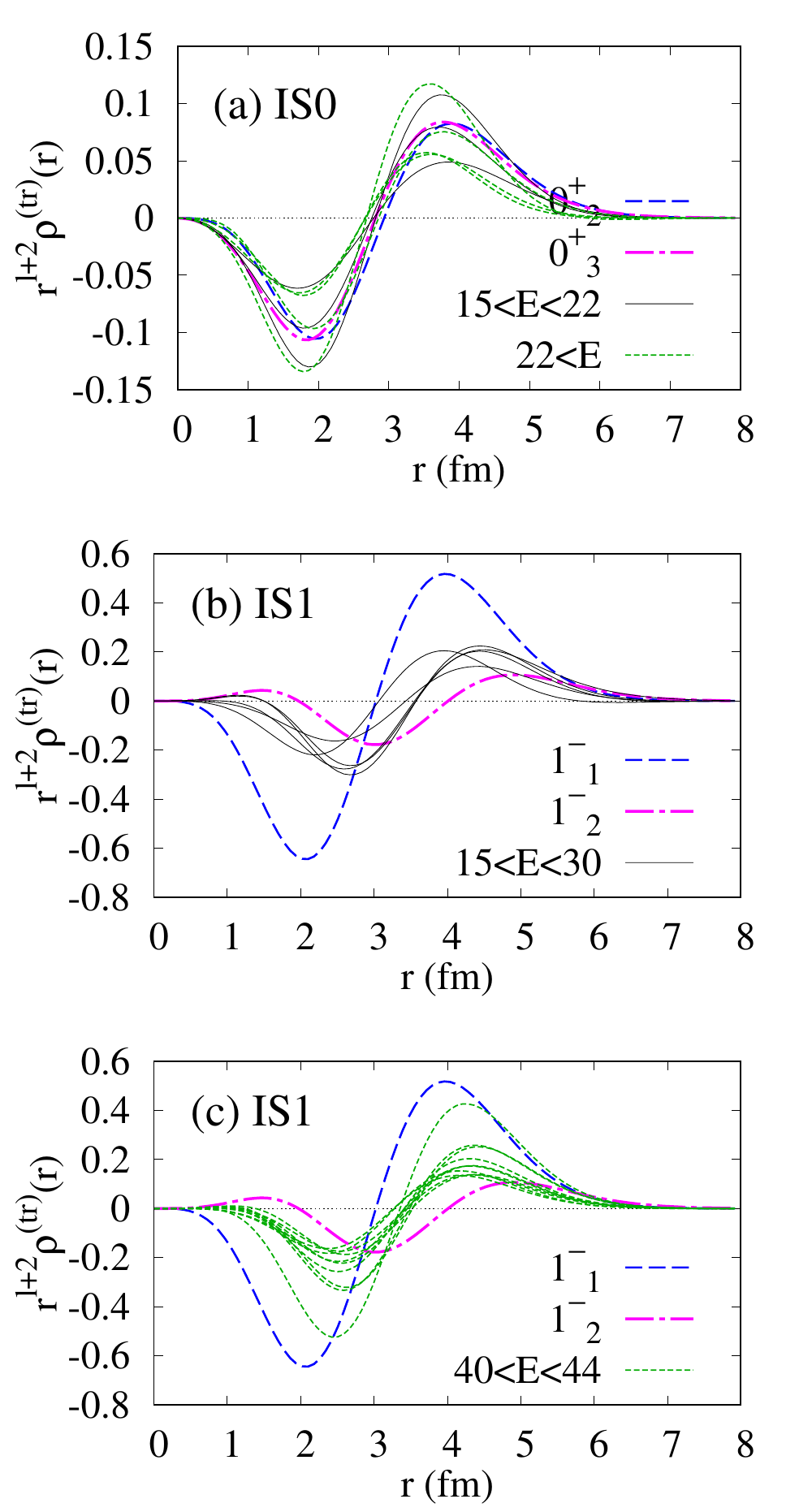} 	
\end{center}
%\vspace{0.5cm}
  \caption{(color online) 
Transition densities of the IS0 and IS1 transitions, 
$0^+_1\to 0^+_k$ and $0^+_1\to 1^-_k$,  
calculated with the sAMD+GCM. 
The densities for the transitions 
with significant strengths as 
$B(\textrm{IS0}:0^+_1\to 0^+_k) \ge 10$ fm$^4$ and 
$B(\textrm{IS1}:0^+_1\to 1^-_k) \ge 10$ fm$^6$ are plotted. 
The transition densities for the first and 
second excited states are shown by blue dashed and magenta dash-dotted lines, respectively. 
The IS0 transition 
density of $0^+$ states in $15<E<22$ MeV and the IS1 transition density of 
$1^-$ states in $15<E<30$ MeV are shown by black solid lines in panels (a) and (b) and those of 
the $0^+$ states in $22<E$ MeV and $1^-$ states in $40 <E <44 $ MeV are shown by 
green dotted lines in panels (a) and (c).
\label{fig:trans}}
\end{figure}

For further discussions of the IS0 and IS1 transitions, we show
the transition densities for the $0^+$ and $1^-$ states with
 $B(\textrm{IS0})> 10$ fm$^4$ and $B(\textrm{IS1})> 10$ fm$^6$
in Fig.~\ref{fig:trans}. The transition density of  the $0^+$ states
in Fig.~\ref{fig:trans}(a) shows qualitatively 
similar behavior with one node around  $r=2.5-3.0$ fm, but one can see a quantitative difference between
the $0^+_2$ and high-energy $0^+$ states. 
The transition density in the $0^+_2$ state is expanded outward 
and its node is located at the largest position $r \sim 3$ fm
due to the developed $^{12}$C$(0^+_1)$+$\alpha$ cluster structure.
Conversely, the transition density of  higher states in  $E> 22~\textrm{MeV}$ is contracted inward. 
This trend can be understood by the character of small amplitude vibration in 
the high-energy monopole excitations. The transition density for other $0^+$ states in $15 <E <22$ MeV shows the
intermediate feature.

Compared with the monopole transitions, the IS1 transition density sensitively reflects different 
characters of dipole excitations. In particular, 
one can see clear differences in the transition density between the $1^-_1$, $1^-_2$, and high-energy GDR.
The transition density in the $1^-_1$ with the compact $4\alpha$ structure shows the most contracted distribution 
with a node at $r<3$ fm and the surface peak at $r\sim 4$ fm. 
On the other hand, in the $1^-_2$ state assigned to the $^{12}$C$+\alpha$ band,  
the transition density has two nodes and shows the broadly stretched distribution with the surface peak at $r\sim 5$ fm.
In the high-energy GDR transition, which are contributed by the $1^-$ states in $40<E<44~\textrm{MeV}$, 
the transition density shows the intermediate feature with one node at $r\sim 3.5$ fm
and the surface peak at $4 \lesssim r \lesssim 4.5~\textrm{fm}$. 
In $15<E<30~\textrm{MeV}$, most of the $1^-$ states have the GDR-like transition density but 
a few states show the $1^-_1$-like contracted behavior. 

%%%%%%%%%%%%%%%%%%%%%%%%%%%%%%
%%%%%%%%%%%%%%%%%%%%%%%%%%%%%

%%%%%%%%%%%%%%%%%%%%%%%%%%%%%%
%%%%%%%%%%%%%%%%%%%%%%%%%%%%%

\subsection{Vortical nature of dipole excitations}

In order to clarify properties of the LED and GDR states, we calculate the transition
strengths with the CD and TD operators. 
Note that the CD strength, which is in principle equivalent to 
the IS1 strength, is sensitive to the compression dipole mode, 
whereas the TD strength can probe the nuclear vorticity in the dipole excitation. 

\begin{figure}[!h]
\begin{center}
\includegraphics[width=8.6cm]{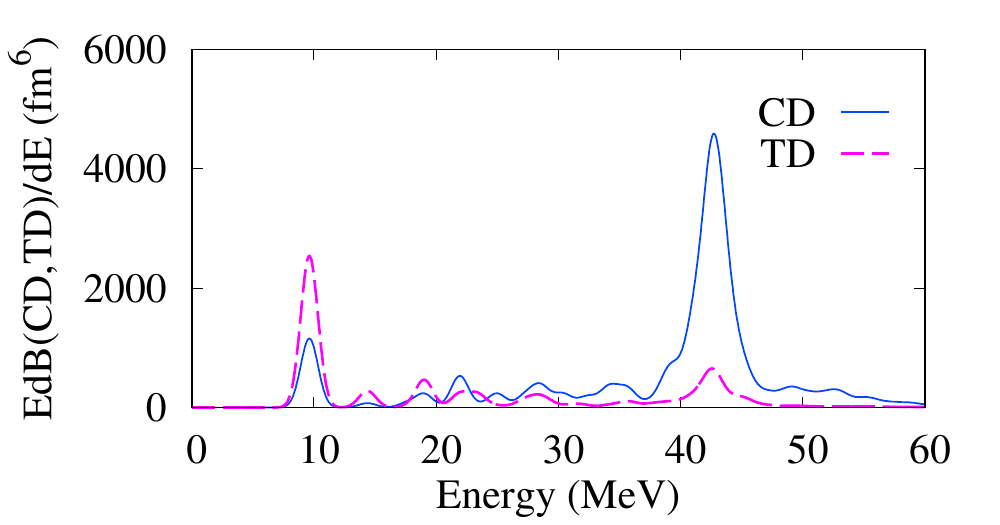} 	
\end{center}
%\vspace{0.5cm}
  \caption{(color online)  
The energy weighted strength functions of the CD and TD transitions calculated with 
the sAMD+GCM. 
The scaled strengths $\tilde B(D)$
of discrete states are smeared by Gaussian with the range 
$\gamma=1/\sqrt{\pi}$ MeV.
\label{fig:cdtm}}
\end{figure}

The calculated CD and TD strength functions
are shown in Fig.~\ref{fig:cdtm}. 
In the CD transitions, we obtain the significant strength below 10 MeV for the $1^-_1$ state and 
the huge peak around $E=40$ MeV for the IS GDR. 
In contract to the CD strength, there is no remarkable TD strength in the high-energy region
for the IS GDR. 
From this result, it is concluded that the IS GDR do not have the vortical feature but is the normal compressive mode. 
Instead, the TD strength is concentrated on the $1^-_1$ state probing the vortical nature.
The $1^-_2$ in the  $^{12}$C+$\alpha$ band has the weak CD and TD transitions
because this state is the  inter-cluster excitation involving the $^{12}$C-cluster rotation and is weakly  
excited by the CD and TD operators. 

The present result indicates quite different characters of the dipole excitations between the $1^-_1$, $1^-_2$, 
and  IS GDR states: the strong CD and TD transitions in the $1^-_1$, weak CD and TD transitions 
in the $1^-_2$, and strong CD but weak TD transitions in the IS GDR. In particular, one of the prominent features of the 
$1^-_1$ is the strong TD strength. In the analysis of the intrinsic wave functions, we find that the TD strength in the 
$0^+_1\to 1^-_1$
is contributed by the dominant $K=1$ component of the prolately deformed $3\alpha$+$\alpha$ structure
of the $1^-_1$. On the other hand, the CD strength in the $0^+_1\to 1^-_1$ is mainly contributed by the 
$K=0$ component. In the $4\alpha$ structure, the $K=1$ and $K=0$ components have large overlap and 
mixes to each other because of the bosonic symmetry of $\alpha$ clusters. This is a unique feature of the 
dipole excitation in $^{16}$O, in which the $1^-_1$ state has the strong TD and CD strengths.

\begin{figure}[!h]
\begin{center}
\includegraphics[width=8.6cm]{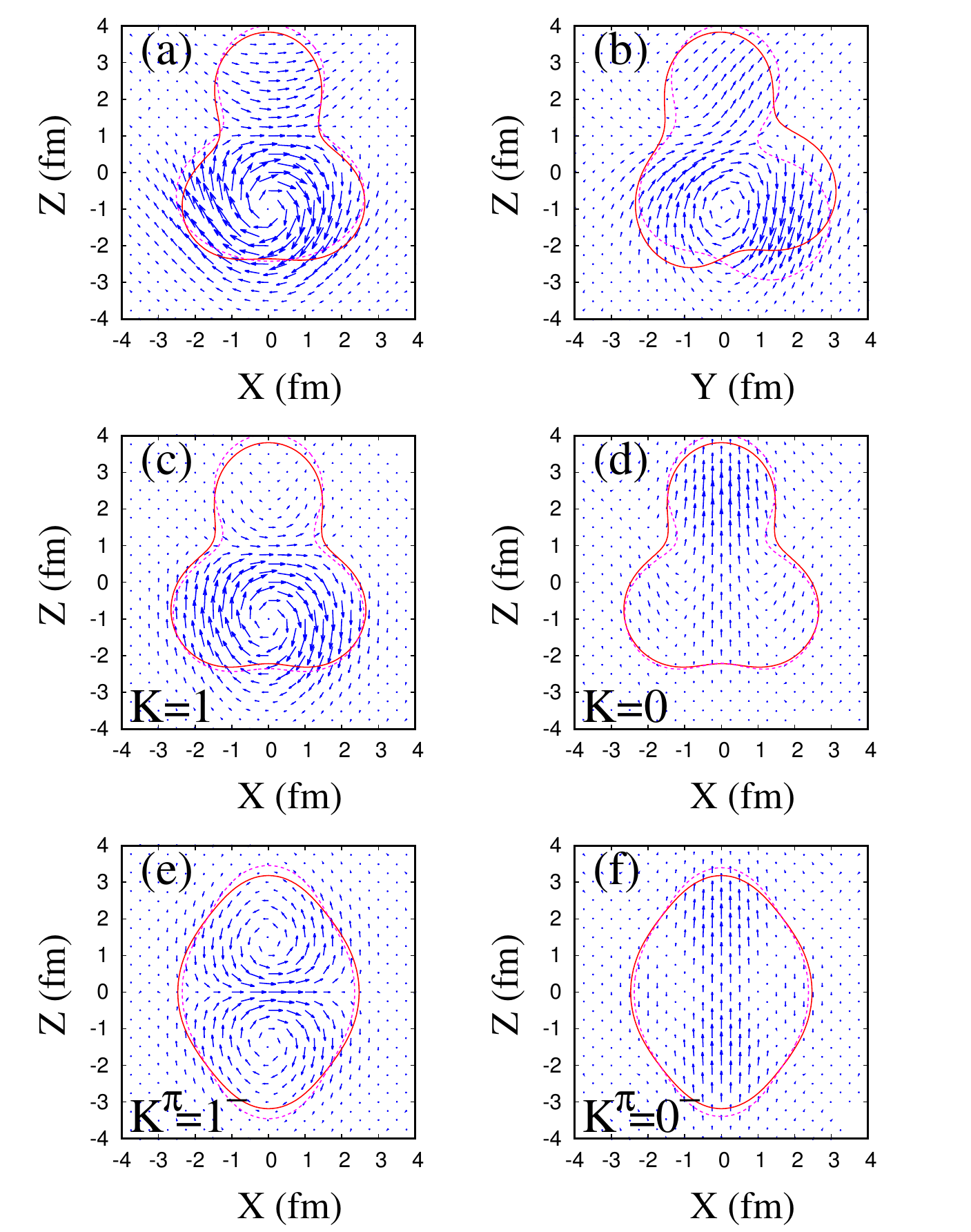} 	
\end{center}
%\vspace{0.5cm}
  \caption{(color online)  
Transition current density of the $0^+_1 \to 1^-_1$in the intrinsic frame
calculated using the wave functions $\Phi^{^{16}\textrm{O}}_\textrm{AMD}(\bvec{Z}^{\rm opt}_{\beta=0^+_1})$ 
and 
$\Phi^{^{16}\textrm{O}}_\textrm{AMD}(\bvec{Z}^{\rm opt}_{\beta=1^-_1})$ 
obtained by the VAP. 
The vector plot of the transition current density before the $K$ and parity projections at the (a) $Y=0$ on the $X$-$Z$ plane
and (b) $X=0$ on the $Y$-$Z$ plane, (c) (d) 
that after the $K$ projection before the parity projection, and  (e) (f)  after the $K$ and parity projections are shown. 
Red solid and magenta dashed lines indicate 
contours for the matter densities 
$\rho(X,0,Z)=0.08$ fm$^{-3}$ of the initial ($0^+_1$) and final $(1^-_1)$ states, respectively.
\label{fig:current}}
\end{figure}

To illustrate the vortical and compressive natures of the $1^-_1$,  we show in Fig.~\ref{fig:current} 
the transition current density of the $0^+_1\to 1^-_1$ transition in the intrinsic frame
calculated using the wave functions $\Phi^{^{16}\textrm{O}}_\textrm{AMD}(\bvec{Z}^{\rm opt}_{\beta=0^+_1})$ 
and 
$\Phi^{^{16}\textrm{O}}_\textrm{AMD}(\bvec{Z}^{\rm opt}_{\beta=1^-_1})$ 
obtained by the VAP. Here, the transition current density before the $K$ and parity projections at the $Y=0$ and $X=0$ planes,
(c) (d) that after the $K$ projection before the parity projection, and (e) (f)  that after the $K$ and parity projections are shown. 
The nuclear matter density of the $0^+_1$ and $1^-_1$ states are also shown by solid and dashed lines, respectively. 
Note that, the parity (axial) symmetry is broken in the intrinsic states before the parity  projection ($K$ projection)
but it is restored after the projection.

In the transition current density before the $K$ and parity projections, a vortex is created at the lower part by the tilting motion  
of the triangle $3\alpha$ in the tetrahedral $4\alpha$ configuration as seen in Fig.~\ref{fig:current}(a) and (b). 
After the $K=1$ projection, where the nuclear current is averaged around the $Z$-axis,  
a $K=1$ vortex appears clearly at the lower part of Fig.~\ref{fig:current}(c). 
Then, after the parity projection, the vortical current is duplicated and 
two vortexes appear in the lower and upper parts.
The $K=1$ vortexes aligned along the prolate deformation is  
the feature of the $K=1$ VD mode 
in the prolately deformed system. This mode  differs from the 
torus-shape vortex, which has been originally proposed in
the $K=0$ dipole excitation
(obviously, the torus current is allowed only in the $K=0$ dipole excitation because of the 
mathematical condition.) 
The geometrical shape of the current in the $K=1$ VD mode is described 
in detail in our previous paper \cite{Shikata:2019wdx}.

Let us turn to the  nuclear current  in the $K=0$ component shown in Fig.~\ref{fig:current}(d) and (f)
before and after the parity projection, respectively.
The $0^+_1\to 1^-_1$ excitation also contains the relative motion between the last $\alpha$ cluster
and the $3\alpha$. In the $K=0$ component, this corresponds to 
the $L=1$ excitation of the $3\alpha$-$\alpha$ relative distance. 
The relative oscillation of the last $\alpha$ cluster against the $3\alpha$ induces the compressive 
nuclear current as seen in Fig.~\ref{fig:current}(d) and (f) and contributes to the
significant CD strength in the $0^+_1\to 1^-_1$ transition.

Strictly speaking, it is not be able to uniquely define the intrinsic frame for physical states 
with eigenvalues of  angular momentum, 
but in the present case that the system has the prolate deformation
because of the tetrahedral $3\alpha+\alpha$ configuration, 
the discussion in the ``intrinsic'' frame can be useful to get the intuitive understanding.

%%%%%%%%%%%%%%%%%%%%%%%%%%%%%%
%%%%%%%%%%%%%%%%%%%%%%%%%%%%%

%%%%%%%%%%%%%%%%%%%%%%%%%%%%%%
%%%%%%%%%%%%%%%%%%%%%%%%%%%%%

\section{Summary and outlook}\label{sec:summary}

The IS monopole and dipole
excitations in $^{16}$O were investigated with the sAMD+GCM. 
The significant  IS0 and IS1 transition strengths were obtained in the low-energy region 
in addition to the GRs.
The $1^-_1$ state contributes to 
the significant low-energy strength of the IS1 transition with 5\% of the energy-weighted sum rule,
which describes well the experimental data observed by $\alpha$ inelastic scattering.
The calculated form factors of the inelastic transitions to the $0^+_2$, $0^+_3$, and $1^-_1$ states 
reproduce the experimental electron scattering form factors. 
The transition densities were also analyzed.

The different characters of the dipole excitations were found in the $1^-_1$, $1^-_2$, 
and  IS GDR: the strong CD and TD transitions in the $1^-_1$, the weak CD and TD transitions 
in the $1^-_2$, and the strong CD but weak TD transitions in the IS GDR. 
Cluster and vortical aspects of the low-energy dipole states were investigated. 
%The $1^-_1$ state is the vortical dipole mode with the tetrahedral $4\alpha$ structure, and 
%the $1^-_2$ state is the $^{12}$C+$\alpha$-cluster state. 
In conclusion, we regard the $1^-_1$ as the vortical vibration mode with the tetrahedral $4\alpha$ structure,  
the $1^-_2$ as the $^{12}$C+$\alpha$ cluster mode, and the IS GDR as the collective vibration of the compressive dipole mode.

%\section*{Acknowledgments} 
\begin{acknowledgments}
The authors would like to thank the Dr.~Nesterenko and Dr.~Chiba for fruitful discussions.
The computational calculations of this work were performed by using the
supercomputer in the Yukawa Institute for theoretical physics, Kyoto University. This work was supported by 
JSPS KAKENHI Grant Nos. 18K03617 and 18J20926. 
\end{acknowledgments}

\appendix

\section{Transition densities}

The density and current density operators for the nuclear matter are defined as  
\begin{eqnarray}
\rho(\bvec{r})&=& \sum_k  \delta(\bvec{r}-\bvec{r}_k),\\
\bvec{j}(\bvec{r})&=& -\frac{i\hbar}{2m} \sum_k  \nabla_k\delta(\bvec{r}-\bvec{r}_k)+\delta(\bvec{r}-\bvec{r}_k)\nabla_k.
\end{eqnarray}  
Here, $\bvec{j}(\bvec{r})$ includes only the convection term of
the nuclear current but not the spin term of magnetization. 
The transition density and current density for the $|0\rangle \to |f\rangle$ transition are given as 
\begin{align}
\rho^\textrm{(tr)}_{0\to f}(\bvec{r})= \langle f |\rho(\bvec{r}) |0 \rangle,\\
\delta\bvec{j}(\bvec{r})= \langle f |\bvec{j}(\bvec{r}) |0 \rangle.
\end{align}  
The $\lambda$th transition density is obtained 
from the multipole decomposition of the transition density, 
\begin{align}
\rho^\textrm{(tr)}_{0\to f}(\bvec{r})&=\frac{1}{\sqrt{2J_f+1}}\sum_\lambda \rho^\textrm{(tr)}_{\lambda;0\to f}(r) \\
&\times
\sum_\mu Y^*_{\lambda\mu}(\hat{\bvec{r}})(J_i M_i \lambda \mu|J_f M_f),
\end{align} 
where $J_i$ and $M_i$  ($J_f$ and $M_f$) are the spin quantum numbers of the initial $|0\rangle$ (final $|f\rangle$) state.
The $\lambda$th multipole component of the so-called longitudinal form factor is related to the Fourier-Bessel transform of the transition 
charge density 
$\rho^\textrm{ch}_{\lambda;0\to f}(r)$ by 
\begin{align}
F(q)=
\frac{\sqrt{4\pi}}{Z}\frac{1}{\sqrt{2J_i+1}} \int dr r^2 j_\lambda(qr)  \rho^\textrm{ch}_{\lambda;0\to f}(r),
\end{align} 
where  $\rho^\textrm{ch}_{\lambda;0\to f}(r)$ is calculated by taking into account the proton charge radius and 
assuming the mirror symmetry.

\section{IS monopole and dipole operators and transition strengths}
The standard compressive-type IS$\lambda$ operators of  the
 IS monopole and dipole excitations are 
defined as 
\begin{align}
M_\textrm{IS0}\equiv&\int d\bvec{r} \rho(\bvec{r}) r^2, \\
M_\textrm{IS1}(\mu)\equiv&\int d\bvec{r} \rho(\bvec{r}) r^3 Y_{1\mu} (\hat{\bvec{r}}).  
\end{align}
The IS0 and IS1 transition strengths for  $|0^+_1\rangle \to |J^\pi_k\rangle$ are given by the 
reduced matrix elements as 
\begin{align}
B(\textrm{IS}\lambda)=\frac{1}{2J_i+1} \left| \langle J^\pi_k||M_ {\textrm{IS}\lambda}||0^+_1 \rangle \right |^2, 
\end{align}
where the angular momentum of the initial state is $J_i$ and that of the final state is 
$J^\pi_k=0^+_k$ and  $1^-_k$ for $\lambda=0$ and 1, respectively. 
The reduced matrix elements are related to 
the transition densities as 
\begin{align}
\langle J^\pi_k||M_ {\textrm{IS}\lambda} ||0^+_1 \rangle = \sqrt{4\pi} \int dr r^2 r^{\lambda+2}  
\rho^\textrm{(tr)}_{\lambda;0\to f}(\bvec{r})
\end{align}
for the IS0 transition and 
\begin{align}
\langle J^\pi_k||M_ {\textrm{IS}\lambda} ||0^+_1 \rangle = \int dr r^2 r^{\lambda+2}  
\rho^\textrm{(tr)}_{\lambda;0\to f}(\bvec{r})
\end{align}
for the IS1 transition. 

The energy-weighted sum rule of the IS0 operator is 
\begin{align}
\sum_k (E_k-E_0) B(\textrm{IS0};0^+_1 \to 0^+_k)= \frac{2\hbar^2 A}{m} \langle r^2 \rangle
\end{align}
with the mean square radius 
$\langle r^2\rangle=\langle 0^+_1| \sum_i r^2_i |0^+_1\rangle/A$ of the ground state.
For the IS1 operator, we use the following energy-weighted sum rule 
from Ref.~\cite{Harakeh:1981zz}, 
\begin{align}
&\sum_k (E_k-E_0) B(\textrm{IS1};0^+_1 \to 1^-_k)& \nonumber\\
&= \frac{3\hbar^2 A}{32 m\pi} \left( 11\langle r^4 \rangle
-\frac{25}{3}\langle r^2\rangle^2 -10 \epsilon \langle r^2\rangle \right),
\end{align}
where $\langle r^4\rangle=\langle 0^+_1| \sum_i r^4_i |0^+_1\rangle/A$ and  
$\epsilon=(4/{\cal E}_2+5/{\cal E}_0)\hbar^2/3mA$. Here ${\cal E}_2$ and 
${\cal E}_0$ are the IS GQR and GMR energies, for which the empirical values of 
${\cal E}_2=63 A^{-1/3}$ MeV
and ${\cal E}_0=80 A^{-1/3}$ MeV are used,  respectively.

\section{CD and TD strengths}

In the analysis of isoscalor dipole excitations,
the CD and TD operators are used as done in Refs.~\cite{Kanada-Enyo:2015knx,Kanada-Enyo:2017fps}.
The former (CD) corresponds to the standard IS1 operator and sensitive to the compressive dipole excitations, 
and the latter (TD) has been proved to be as a good measure of the nuclear vorticity in the dipole excitations
as discussed in Ref.~\cite{Kvasil:2011yk}. 
They are defined as 
\begin{eqnarray}
M_\textrm{CD}(\mu)&=&\frac{-i}{2\sqrt{3}c}\int d\bvec{r} \bvec{j}(\bvec{r}) 
 \nonumber\\
&\cdot& 
\left [  \frac{2\sqrt{2}}{5} r^2 \bvec{Y}_{12\mu}(\hat{\bvec{r}}) - r^2 \bvec{Y}_{10\mu} (\hat{\bvec{r}}) 
\right ],\\
M_\textrm{TD}(\mu)&=&\frac{-i}{2\sqrt{3}c}\int d\bvec{r} \bvec{j}(\bvec{r}) \nonumber\\
&\cdot& 
\left [
\frac{\sqrt{2}}{5} r^2 
\bvec{Y}_{12\mu}(\hat{\bvec{r}})+
r^2 \bvec{Y}_{10\mu} (\hat{\bvec{r}})  
\right ],
\end{eqnarray}
where $\bvec{Y}_{\lambda L\mu}$ is the vector spherical  harmonics.

The matrix elements of these IS dipole operators for the $|0^+_1\rangle \to |1^-_k\rangle$ transitions
are given 
as 
\begin{eqnarray}
&&\langle 1^-_k|M_\textrm{CD}(\mu) |0^+_1  \rangle=\nonumber\\
&&\frac{-i}{2\sqrt{3}c}\int d\bvec{r} \delta\bvec{j}(\bvec{r}) \cdot
\left [  \frac{2\sqrt{2}}{5} r^2 \bvec{Y}_{12\mu}(\hat{\bvec{r}}) - r^2 \bvec{Y}_{10\mu} (\hat{\bvec{r}}) 
\right ],\nonumber\\
\ \\
&&\langle  1^-_k|M_\textrm{TD}(\mu) |0^+_1  \rangle =\nonumber\\
&&\frac{-i}{2\sqrt{3}c}\int d\bvec{r} \delta\bvec{j}(\bvec{r})\cdot
\left [
\frac{\sqrt{2}}{5} r^2 
\bvec{Y}_{12\mu}(\hat{\bvec{r}})+
r^2 \bvec{Y}_{10\mu} (\hat{\bvec{r}})  
\right].\nonumber\\
\ 
\end{eqnarray}
Using the continuity equation, 
the CD matrix element is related to 
the matrix element of the standard IS1 operator $M_\textrm{IS1}$ as 
\begin{eqnarray}
\langle 1^-_k|M_\textrm{CD}(\mu)|  0^+_1 \rangle &=&-
\frac{E}{10\hbar c} \langle 1^-_k |M_\textrm{IS1}(\mu)|  0^+_1\rangle.
\end{eqnarray}
The CD and TD strengths, which are scaled with the 
factor  $\left(\frac{10\hbar c}{E}\right )^2$, are defined as 
\begin{eqnarray}
&&\tilde B(\textrm{CD,TD}; 0^+_1 \to 1^-_k) \nonumber\equiv  \left(\frac{10\hbar c}{E}\right )^2 \left| \langle 1^-_k||M_\textrm{CD,TD} ||0^+_1 \rangle \right |^2.\\
\end{eqnarray}

\end{document}